# Applying Fuzzy ID3 Decision Tree for Software Effort Estimation

Ali Idri[1] and Sanaa Elyassami[1]

[1] Department of Software Engineering

ENSIAS, Mohammed Vth –Souissi University,

BP. 713, Madinat Al Irfane, Rabat, Morocco

**Abstract**

Web Effort Estimation is a process of predicting the efforts and cost in terms of money, schedule and staff for any software project system. Many estimation models have been proposed over the last three decades and it is believed that it is a must for the purpose of: Budgeting, risk analysis, project planning and control, and project improvement investment analysis. In this paper, we investigate the use of Fuzzy ID3 decision tree for software cost estimation; it is designed by integrating the principles of ID3 decision tree and the fuzzy set-theoretic concepts, enabling the model to handle uncertain and imprecise data when describing the software projects, which can improve greatly the accuracy of obtained estimates. MMRE and Pred are used as measures of prediction accuracy for this study. A series of experiments is reported using two different software projects datasets namely, Tukutuku and COCOMO'81 datasets. The results are compared with those produced by the crisp version of the ID3 decision tree.

## 1. Introduction

Estimation software project development effort remains a complex problem, and one which continues to attract considerable research attention. Improving the accuracy of the effort estimation models available to project managers would facilitate more effective control of time and budgets during software project development. Unfortunately, many software development estimates are quite inaccurate. Molokken and Jorgensen report in recent review of estimation studies that software projects expend on average 30-40% more effort than is estimated [13]. In order to make accurate estimates and avoid gross misestimations, several cost estimation techniques have been developed. These techniques may be grouped into two major categories: parametric models, which are derived from the statistical or numerical analysis of historical projects data [5], and non-parametric models, which are based on a set of artificial intelligence techniques such as artificial neural networks [9][4], case based reasoning [19], decision trees [20] and fuzzy logic [22][17]. In this paper, we are concerned with cost estimation models based on fuzzy decision trees especially Fuzzy Interactive Dichotomizer 3.

The decision tree method is widely used for inductive learning and has been demonstrating its superiority in terms of predictive accuracy in many fields [23][10]. The most widely used algorithms for building a decision tree are ID3 [11], C4.5 [12] and CART [14].

There are three major advantages when using estimation by decision trees (DT). First, decision trees approach may be considered as "white boxes", it is simple to understand and easy to explain its process to the users, contrary to other learning methods. Second, it allows the learning from previous situations and outcomes. The learning criterion is very important for cost estimation models because software development technology is supposed to be continuously evolving. Third, it may be used to feature subset selection to avoid the problem of cost driver selection in software cost estimation model.

On the other hand, fuzzy logic has been used in software effort estimation. It's based on fuzzy set theory, which was introduced by Zadeh in 1965 [15]. Attempts have been made to rehabilitate some of the existing models in order to handle uncertainties and imprecision problems. Idri et al. [3] investigated the application of fuzzy logic to the cost drivers of intermediate COCOMO model while Pedrycz et al. [24] presented a fuzzy set approach to effort estimation of software projects.

In two earlier works [1][2] we have empirically evaluated the use of crisp decision tree techniques for software cost estimation. More especially, the two used crisp decision tree techniques are the ID3 and the C4.5 algorithms. The two studies are based on the COCOMO' 81 and a web hypermedia dataset. We have found that the decision tree designed with the ID3 algorithm performs better, in terms of cost estimates accuracy, than the decision tree designed with C4.5 algorithm for the two datasets.

The aim of this study is to evaluate and to discuss the use of fuzzy decision trees, especially the fuzzy ID3 algorithm in designing DT for software cost estimation. Instead of crisp DT, fuzzy DT may allow to exploit complementary advantages of fuzzy logic theory which is the ability to deal with inexact and uncertain information when describing the software projects.

The remainder of this paper is organised as follows: In section II, we present the fuzzy ID3 decision tree for software cost estimation. The description of datasets used to perform the empirical studies and the evaluation criteria adopted to measure the predictive accuracy of the designed models are given in section III. Section IV focuses on the experimental design. In Section V, we present and discuss the obtained results when the fuzzy ID3 is used to estimate the software development effort. A comparison of the estimation results produced by means of the fuzzy ID3 models and the crisp ID3 model is also provided in section and section V. A conclusion and an overview of future work conclude this paper.

## 2. Fuzzy ID3 for Software Cost Estimation

Based on the Concept Learning System algorithm, Quinlan proposed a decision tree called the Interactive Dichotomizer 3 (ID3). The ID3 technique is based on information theory and attempts to minimize the expected number of comparisons.

The fuzzy ID3 is based on a fuzzy implementation of the ID3 algorithm [16][21]. It's formed of one root node, which is the tree top, or starting point, and a series of other nodes. Terminal nodes are leaves (effort). Each node corresponds to a split on the values of one input variable (cost drivers). This variable is chosen in order to reach a maximum of homogeneity amongst the examples that belong to the node, relatively to the output variable.

Figure 1 illustrates an example of fuzzy ID3 decision tree for software development effort where MF represents the membership function used to define fuzzy sets for each cost driver.

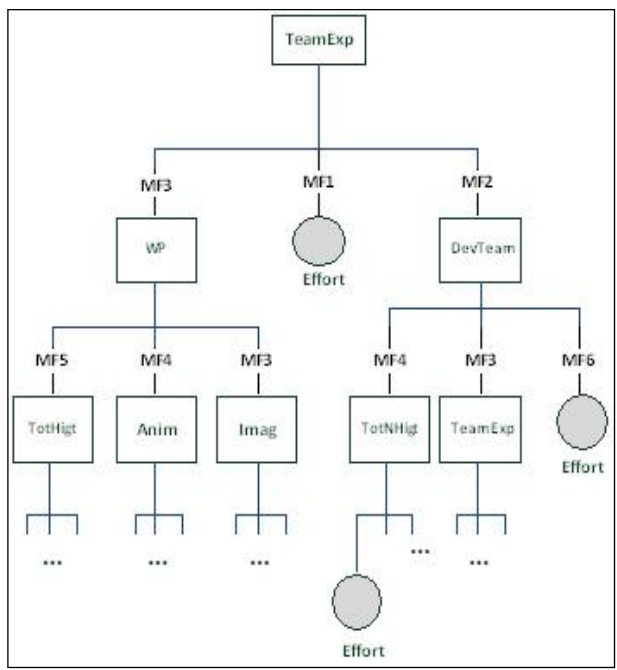

Fig. 1 An example of fuzzy ID3 decision tree for software development effort

The major characteristic of fuzzy ID3 is that an example belongs to a node to a certain degree. The proportion of $p_k^n$ of examples with $k$ classification at node $n$ is calculated using the membership degrees as follows:

$$p_k^n = \frac{\sum_{i=1}^{N} u_k(y_i) \wedge u_n(x_i)}{\sum_{c=1}^{K} \sum_{i=1}^{N} u_c(y_i) \wedge u_n(x_i)} \quad (1)$$

Where K represents the classes and N is the number of examples in the subset. $u_k(y_i)$ is the membership degree on the web project $i$ that belongs to the class $k$ and $u_n(x_i)$ is the membership degree of the web project $i$ at node $n$.

$\wedge$ represents the conjunction operator. T-norm, which generalizes intersection in the domain of fuzzy sets, is usually used for fuzzy conjunction. The most popular T-norms are minimum and product.

The fuzzy entropy uses the membership degree of examples at a particular node and contributes to enhance the discriminative power of an attribute, is computed as:

$$H_n = -\sum_k p_k^n * \log(p_k^n) \quad (2)$$

The growth of the fuzzy ID3 is realized by expanding a node of tree characterized by the highest information gain. The information gain is calculated as follows:

$$G_n^j = H_n - \sum_{l=1}^{M} w_l H_l \quad (3)$$

Where $H_n$ is the entropy in the node $n$. $H_l$ is the entropy of the node that belongs to the fuzzy set $L$ of the $j$ variable. $w_l$ is the fuzzy set relative weight.

The node $n$ is split into as many sub-nodes as there are attributes. The algorithm terminated when all attributes are used for splits, or when all examples at a node have the same classification.

## 3. Data Description and Evaluation Criteria

This section describes the dataset used to perform this empirical study and the evaluation criteria adopted to measure the estimates accuracy of the designed software cost estimation model based on fuzzy ID3 method.

3.1 Data Description

In this empirical study, two historical software projects datasets are used:
        1- Tukutuku dataset [7]
        2- COCOMO'81 dataset [5]

**The Tukutuku dataset** contains 53 web projects. Each web application is described using 9 numerical attributes such as: the number of html or shtml files used, the number of media files and team experience (see Table I).

However, each project volunteered to the Tukutuku database was initially characterized using more than 9 software attributes, but some of them were grouped together. For example, we grouped together the following three attributes: the number of new Web pages developed by the team, the number of Web pages provided by the customer and the number of Web pages developed by a third party (outsourced) in one attribute reflecting the total number of Web pages in the application (TotWP).

Table I: Software Attributes for the Tukutuku dataset

| Attributes | Description |
|---|---|
| TeamExp | Average team experience with the development language(s) employed |
| DevTeam | Size of development team |
| TotWP | Total number of web pages |
| TextPages | Number text pages typed (~600 words) |
| TotImg | Total number of images |
| Anim | Number of animations |
| AV | Number of audio/video files |
| TotHigh | Total Number of high effort features/functions |
| TotNHigh | Total Number of low effort features/functions |

**The COCOMO'81 dataset** contains 252 software projects which are mostly scientific applications developed by Fortran. Each software project is described using 13 attributes: software size measured in KDSI (Thousands of Delivered Source Instructions) and the remaining 12 numerical attributes described in Table II.

The 12 numerical attributes describe the environment in which the program will be designed to operate, the relationship between a program and its host or developmental platform, selected project management facets of a program such as the experience of the personnel involved in the software project, the time and storage constraints imposed on the software and the method used in the development.

Table II: Software Attributes for the COCOMO'81 dataset

| Attributes | Description |
|---|---|
| SIZE | Software Size |
| DATA | Database Size |
| VIRTMIN, VIRTMAJ | Virtual Machine Volatility |
| TIME | Execution Time Constraint |
| STOR | Main Storage Constraint |
| TURN | Computer Turnaround Time |
| ACAP | Analyst Capability |
| AEXP | Applications Experience |
| PCAP | Programmer Capability |
| VEXP | Virtual Machine Experience |
| LEXP | Programming Language Experience |
| SCED | Required Development |

3.2 Evaluation criteria

We employ the following criteria to measure the accuracy of the estimates generated by the fuzzy ID3. A common criterion for the evaluation of effort estimation models is the magnitude of relative error (MRE), witch is defined as

$$MRE = \left| \frac{Effort_{actual} - Effort_{estimated}}{Effort_{actual}} \right| \quad (4)$$

where $Effort_{actual}$ is the actual effort of a project in the dataset, and $Effort_{estimated}$ is the estimated effort that was obtained using a model or a technique.

The *MRE* values are calculated for each project in the datasets, while mean magnitude of relative error (MMRE) computes the average over *N* projects

$$MMRE = \frac{1}{N}\sum_{i=1}^{N}\left|\frac{Effort_{actual,i} - Effort_{estimated,i}}{Effort_{actual,i}}\right| \times 100 \quad (5)$$

The acceptable target values for *MMRE* are $MMRE \leq 25$. This indicates that on the average, the accuracy of the established estimation model would be less than 25%.

Another widely used criterion is the prediction *Pred(p)* witch represents the percentage of *MRE* that is less than or equal to the value p among all projects. This measure is often used in the literature and is the proportion of the projects for a given level accuracy [18]. The definition of *Pred(p)* is given as follows:

$$Pred(p) = \frac{k}{N} \quad (6)$$

Where *N* is the total number of observations and *k* is the number of observations whose *MRE* is less or equal to *p*. A common value for *p* is 25, witch also used in the present study. The prediction at 25%, *Pred(25)*, represents the percentage of projects whose *MRE* is less or equal to 25%. The acceptable values for *Pred(25)* are $Pred(25) \geq 75$.

## 4. Experiment Design

This section describes the experiment design of the fuzzy ID3 decision tree on the both Tukutuku and COCOMO'81 datasets.

The use of fuzzy ID3 to estimate software development effort requires the determination of the parameters, namely the number of input variables, the maximum number of fuzzy sets for each input variable, the significant level value and the conjunction operator. The last two parameters play an essential role in the generation of Fuzzy Decision trees. It greatly affects the calculation of fuzzy entropy and classification results of Fuzzy Decision trees.

The number of input variables is the number of the attributes describing the historical software projects in the used dataset. Therefore, when applying fuzzy ID3 to Tukutuku dataset, the number of input variables is equal to 9 and is equal to 13 in the case of COCOMO'81 dataset. Concerning the maximum number of fuzzy sets is the maximum partition size for each variable, is fixed to 7 for all experiments.

In the present paper we are interested in studying the impact of the fuzzy conjunction operators (t-norms) and the significant level parameter (β) on the accuracy of fuzzy ID3. The significant level is the membership degree for an example to be considered as belonging to the node.

For each dataset, two models of fuzzy ID3 were generated. The first Fuzzy ID3 effort estimation model uses the product entropy conjunction operator to measure the fuzzy entropy (t-norm=product), and the second model uses the minimum entropy conjunction operator to calculate the fuzzy entropy (t-norm=min). These conjunction operators are the two commonly used t-norm operators because of their well behaviour and their computational simplicity [6].

The minimum entropy conjunction operator is defined as:

$$u_k(y_i) \wedge u_n(x_i) = \min[u_k(y_i), u_n(x_i)] \quad (7)$$

Concerning, the product entropy conjunction operator is given as:

$$u_k(y_i) \wedge u_n(x_i) = u_k(y_i) * u_n(x_i) \quad (8)$$

For each model, a series of experiments is conducted with the fuzzy ID3 algorithm each time using a different value of the significant level parameter (β). The significant level is varied within the interval [0, 1].

## 5. Overview of the Empirical Results

This section presents and discusses the results obtained when applying the fuzzy ID3 to the Tukutuku and COCOMO'81 datasets. The calculations were made using Fispro software [8]. We conducted several experiments using different configurations of fuzzy ID3. For these experiments, a holdout validation on the entire datasets was performed. Datasets were randomly split into two groups: training set and test set.

5.1 Tukutuku dataset

The first experiment is performed using Tukutuku dataset containing 53 historical software projects. Two models of fuzzy ID3 were designed. The first Fuzzy ID3 effort estimation model (Model 1) uses the formula of the conjunction operator given in Eq. (8) to compute fuzzy entropy, and the second model (Model 2) uses the formula of the conjunction operator given in Eq. (7). For each model, different configurations have been obtained by varying the significant level (β). The aim is to determine which configuration improves the estimates.

We have trained and tested the two models using Tukutuku dataset. The results for the different configurations have been compared. Figure 2 and figure 3 show the accuracy of the two fuzzy ID3 models, measured in terms of MMRE and Pred, on Tukutuku dataset.

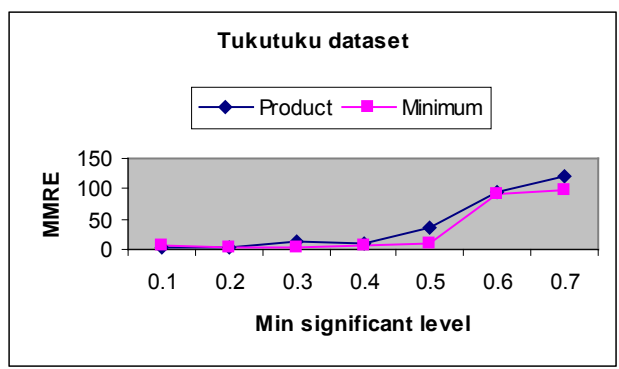

Fig. 2 Relationship between the accuracy of Fuzzy ID3 (MMRE), the used conjunction operator and the SL value

Figure 2 compares the accuracy of the two models, in terms of MMRE, when varying the significant level. We note that the fuzzy ID3 model using the product conjunction operator generates a lower MMRE that the other model using the minimum conjunction operator for significant level value less than 0.2.

For example, for β=0.1 the model 1 generates a lower prediction error (MMRE=2.45) than the model 2 (MMRE=5.31). By against, model 2 generates a lower MMRE than the model 1 for significant level value greater than or equal to 0.2. For example, for β=0.5 the model 2 generate a lower prediction error (MMRE=9.09) than the model 1 (MMRE=34.48).

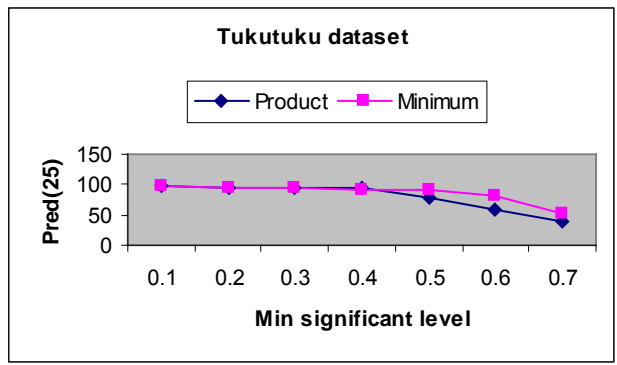

Fig. 3 Relationship between the accuracy of Fuzzy ID3 (Pred), the used conjunction operator and the SL value

Figure 3 shows and compares the results of the two models, in terms of Pred(25), when varying the significant level. From this figure, we note that the accuracy of fuzzy ID3 model using minimum conjunction operator performs much better than fuzzy ID3 model using product conjunction operator for significant level value greater than or equal to 0.2. So, model 2 generates acceptable effort estimates with significant level value less or equal to 0.6. We note that the accuracy of fuzzy ID3 model using the minimum conjunction operator performs much better than fuzzy ID3 model using the product conjunction operator for almost every value of significant level.

Table III summarizes the results obtained using different configurations of fuzzy ID3 for Tukutuku dataset. It shows the variation of the accuracy according to the significant level value and to the used conjunction operator.

Table III: MMRE and Pred results of different fuzzy ID3 configurations for Tukutuku dataset

| Significant level (β) | Accuracy of Fuzzy ID3 | | | |
|---|---|---|---|---|
| | T-norm = Product | | T-norm = Minimum | |
| | MMRE | Pred(25) | MMRE | Pred(25) |
| 0.1 | 2,45 | 97,73 | 5,31 | 97,73 |
| 0.2 | 4,09 | 95,45 | 1,82 | 93,18 |
| 0.3 | 11,7 | 93,18 | 3,87 | 95,45 |
| 0.4 | 8,49 | 93,18 | 5,82 | 90,91 |
| 0.5 | 34,48 | 77,27 | 9,09 | 90,91 |
| 0.6 | 93,08 | 58,49 | 90 | 83,02 |
| 0.7 | 119,3 | 39,62 | 97,41 | 50,94 |
| 0.8 | 210,66 | 26,42 | 111,83 | 45,28 |
| 0.9 | 176,99 | 22,64 | 176,99 | 20,75 |

## 5.2 COCOMO'81 dataset

In the second experiment, we have replicated the previous empirical study using COCOMO'81 dataset to verify how much the use of an adequate conjunction operator affects the accuracy of fuzzy ID3 in estimating software effort.

We have conducted several experiments on the two models. To compute fuzzy entropy, model 1 u ses the product conjunction operator. On the other side, model 2 uses the minimum conjunction operator for calculating fuzzy entropy.

For each model, we varied the significant level (β) from 0.1 to 0.9 degree. Figure 3 and figure 4 show the accuracy of the two fuzzy ID3 models, measured in terms of MMRE and Pred, on COCOMO'81 dataset.

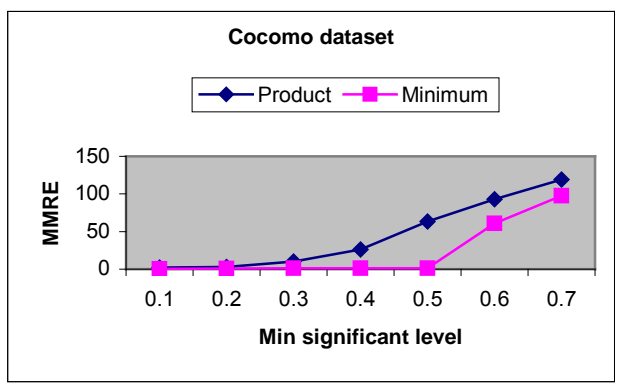

Fig. 4 Relationship between the accuracy of Fuzzy ID3 (MMRE), the used conjunction operator and the SL value

Figure 4 compares the accuracy of the two models, in terms of MMRE, when varying the significant level. From this figure, we note that the accuracy of fuzzy ID3 model using the minimum conjunction operator performs much better than fuzzy ID3 model using the product conjunction operator for each value of significant level. Therefore, in terms of MMRE, Model 2 performs better than Model 1.

Figure 5 shows the results of the two models, in terms of Pred(25), when varying the significant level. From these figures, we confirm the superiority of the fuzzy ID3 model using Eq. (7) to compute the fuzzy entropy over that one using Eq. (8).

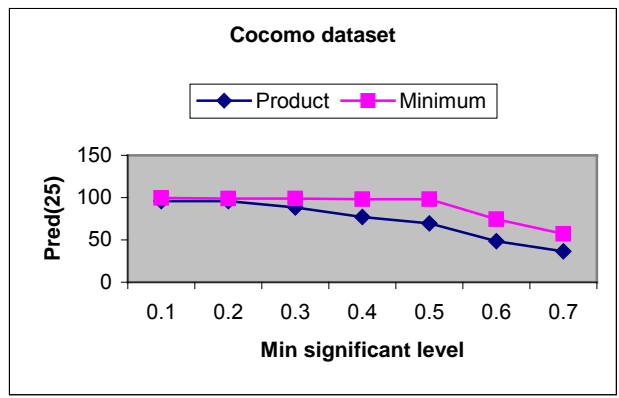

Fig. 5 Relationship between the accuracy of Fuzzy ID3 (Pred) the used conjunction operator and the SL value

Table IV summarizes the results obtained using different configurations of fuzzy ID3 for COCOMO'81 dataset. It shows the variation of the accuracy according to the significant level value and to the used conjunction operator.

Table IV: MMRE and Pred results of different fuzzy ID3 configurations for COCOMO'81 dataset

| Significant level (β) | Accuracy of Fuzzy ID3 | | | |
|---|---|---|---|---|
| | T-norm = Product | | T-norm = Minimum | |
| | MMRE | Pred(25) | MMRE | Pred(25) |
| 0.1 | 1,98 | 95,93 | 0,56 | 99,5 |
| 0.2 | 2,84 | 95,93 | 0,98 | 98,94 |
| 0.3 | 10,11 | 88,24 | 1,17 | 98,94 |
| 0.4 | 26,16 | 76,92 | 1,21 | 98,19 |
| 0.5 | 63,31 | 69,68 | 1,34 | 98,19 |
| 0.6 | 93,08 | 48,41 | 60,86 | 74,6 |
| 0.7 | 119,89 | 36,51 | 97,41 | 57,14 |
| 0.8 | 123,25 | 22,62 | 111,83 | 31,75 |
| 0.9 | 127,32 | 15,08 | 176,99 | 21,43 |

### 3.3 Comparisons between crisp and fuzzy ID3

The comparisons between the results produced by the two fuzzy ID3 models used in the latest subsections (A and B) and the crisp version of ID3 decision tree are shown in table V. For the Tukutuku dataset, the crisp ID3 model generated in [1] is used for the comparison and in the case of the COCOMO'81 dataset, we used the crisp ID3 model applied in [2]. The best results obtained by means of the three models are compared in terms of MMRE and Pred(25).

Table V: Result of the different models used on COCOMO'81 and on Tukutuku datasets

| Performance Criteria | COCOMO'81 dataset | | |
|---|---|---|---|
| | Crisp ID3 | Model 1 | Model 2 |
| MMRE | 28 | 1,98 | 0,56 |
| Pred(25) | 84 | 95,93 | 99,5 |
| Performance Criteria | Tukutuku dataset | | |
| | Crisp ID3 | Model 1 | Model 2 |
| MMRE | 24 | 2,45 | 1,82 |
| Pred(25) | 96 | 97,93 | 97,93 |

The experimental results show that the fuzzy ID3 models show better estimation accuracy than the crisp ID3 model in terms of MMRE and Pred(25). For example, in the case of the COCOMO'81 dataset, the improvement is 92% based on the model 1 MMRE and the crisp ID3 MMRE and is the 98% based on the model 2 MMRE and the crisp ID3 MMRE.

## 4. Conclusions

In this paper, we have empirically studied two fuzzy ID3 models for software effort estimation. Each one used a different formula to compute the fuzzy entropy. These fuzzy ID3 models were trained and tested using two software projects datasets. The results show that the use of an optimal significant level value and an adequate conjunction operator for computing the fuzzy entropy improves greatly the estimates generated by fuzzy ID3 model. The comparison with the crisp version of ID3 decision tree shows encouraging results.

**A. Idri** is a Professor at Computer Science and Systems Analysis School (ENSIAS, Rabat, Morocco). He received DEA (Master) (1994) and Doctorate of 3rd Cycle (1997) degrees in Computer Science, both from the University Mohamed V of Rabat. He has received his Ph.D. (2003) in Cognitive Computer Sciences from ETS, University of Quebec at Montreal. His research interests include software cost estimation, software metrics, fuzzy logic, neural networks, genetic algorithms and information sciences.

**S. Elyassami** received her engineering degree in Computer Science from the UTBM, Belfort-Montbeliard, France, in 2006. Currently, she is preparing her Ph.D. in computer science in ENSIAS. Her research interests include software cost estimation, software metrics, fuzzy logic and decision trees.